\documentclass[12pt]{article}
\usepackage{epsfig}
\usepackage{amssymb}
\usepackage{amsmath}
\usepackage{amsfonts}
\usepackage{graphicx}
\usepackage{mathrsfs}
\usepackage[dvips]{color}
\usepackage{multirow}

\newcommand{\R}{\mathbb{R}}
\newcommand{\C}{\mathbb{C}}

\newcommand{\fa}{\mathfrak{a}}

\newcommand{\fz}{\mathfrak{z}}

\newcommand{\bI}{\mathbf{I}}

\newcommand{\bM}{\mathbf{M}}

\newcommand{\cC}{\mathcal{C}}

\newcommand{\cO}{\mathcal{O}}

\newcommand{\cS}{\mathcal{S}}
\newcommand{\cT}{\mathcal{T}}
\newcommand{\cU}{\mathcal{U}}

\newcommand{\be}{\begin{equation}}
\newcommand{\ee}{\end{equation}}
\newcommand{\bea}{\begin{eqnarray}}
\newcommand{\eea}{\end{eqnarray}}
\newcommand{\nn}{\nonumber}

\newcommand{\ed}{\end{document}}

\newcommand{\bi}{\begin{itemize}}
\newcommand{\ei}{\end{itemize}}

\newcommand{\bce}{\begin{center}}
\newcommand{\ece}{\end{center}}

\newcommand{\sH}{\mathscr{H}}

\newcommand{\sU}{\mathscr{U}}

\newcommand{\IM}{{\rm Im}}

\begin{document}

\title{Dynamical Theory of Scattering, Exact Unidirectional Invisibility, and Truncated $\fz\,e^{-2ik_0x}$ potential}

\author{Ali~Mostafazadeh\thanks{E-mail address: amostafazadeh@ku.edu.tr}~
\\ Departments of Mathematics 
and Physics, 
Ko\c{c} University,\\ 34450 Sar{\i}yer,
Istanbul, Turkey}

\date{ }
\maketitle

\begin{abstract}

The dynamical formulation of time-independent scattering theory that is developed in [Ann.\ Phys.\  (NY) \textbf{341}, 77-85 (2014)] offers simple formulas for the reflection and transmission amplitudes of finite-range potentials in terms of the solution of an initial-value differential equation. We prove a theorem that simplifies the application of this result and use it to give a complete characterization of the invisible configurations of the truncated $\fz\,e^{-2ik_0 x}$ potential to a closed interval, $[0,L]$, with $k_0$ being a positive integer multiple of $\pi/L$. This reveals a large class of exact unidirectionally and bidirectionally invisible configurations of this potential. The former arise for particular values of $\fz$ that are given by certain zeros of Bessel functions. The latter occur when the wavenumber $k$ is an integer multiple of $\pi/L$ but not of $k_0$. We discuss the optical realizations of these configurations and explore spectral singularities of this potential.
\vspace{2mm}

 \vspace{2mm}

\end{abstract}

\section{Introduction}

The idea that a scattering potential can be visible from one direction and invisible from the other is a mind-bugling possibility that has attracted the attention of a large number of researchers in recent years \cite{invisible-1,lin,longhi-jpa-2011,invisible-2,jones-jpa-2012,invisible-3,pra-2013a,ap-2014,pra-2014a,pra-2014b,invisible-4,
pra-2015a,pra-2015b}. Although a particular example of these so-called unidirectionally invisible potentials was examined in Refs.~\cite{invisible-1}, it was the publication of Ref.~\cite{lin} that made the subject into an active area of research. The work done in \cite{invisible-1,lin} focuses on the finite-range potential:
    \be
    v(x)=\left\{\begin{array}{ccc}
    \fz\, e^{-2ik_0x} & {\rm for} & x\in [0,L],\\
    0 & {\rm for} & x\notin [0,L],\end{array}\right.
    \label{exp-pot}
    \ee
where $\fz$ is a real or complex coupling constant, and $k_0$ and $L$ are nonzero real parameters. It turns out that for an incident wave with wavenumber $k=k_0$, this potential is unidirectionally invisible from the right provided that $k_0=\pi/L$. This observation, which was originally made within the confines of a rotating wave approximation \cite{lin}, was later examined more thoroughly in \cite{longhi-jpa-2011,jones-jpa-2012}. The exact results confirmed the validity of the rotating wave approximation for the physically relevant values of the parameters of the system, where $|\fz|/k_0^2\ll 1$, but revealed small discrepancies showing that the unidirectional invisibility property of (\ref{exp-pot}) holds only approximately \cite{longhi-jpa-2011,jones-jpa-2012}. It was subsequently realized that this property is merely a first-order perturbative effect \cite{pra-2014a}.

We can identify (\ref{exp-pot}) as an optical potential modeling the interaction of an infinite planar slab of optical material with normally-incident transverse electric waves provided that we express it in the form
	\be
	v(x)=k^2[1-\varepsilon(x)],
	\label{coupling}
	\ee
where $k$ is the wavenumber of the incident wave that propagates along the $x$-axis, and $\varepsilon(x)$ is the complex permittivity of the slab.  Stacking $m$ copies of such a slab, which is equivalent to taking $k_0=m\pi/L$, one expects to find a potential with the same unidirectional invisibility property. As initially noted in \cite{longhi-jpa-2011}, for sufficiently large values of $m$, this is not the case. We can simply attribute this observation to the fact that this property of (\ref{exp-pot}) holds only perturbatively \cite{pra-2014a}. For sufficiently large values of $m$, the higher order terms in the perturbative expansion of the scattering data (in powers of $\fz$) produce sizable contributions which can no longer be neglected. Indeed, taking into account the second- and third-order terms lead to the violation of the transparency (deviation of the transmission amplitude from unity) and right-reflectionlessness of (\ref{exp-pot}),  respectively \cite{pra-2014a}. This in turn explains the findings of  \cite{longhi-jpa-2011} pertaining the existence of the unidirectional invisible, unidirectional reflectionless, and bidirectional reflectionfull regimes for this potential.

A remarkable feature of the unidirectionally invisible potentials is their role in a local (single-mode) inverse scattering procedure that allows for the construction of a finite-range potential with arbitrary pre-determined scattering properties at a given wavenumber as the sum of up to four finite-range unidirectionally invisible potentials \cite{pra-2014b}. This procedure has interesting applications in devising various unidirectional and bidirectional amplifiers, absorbers, phase shifters \cite{pra-2014b}, and invisibility cloaks \cite{pra-2015b}. Its implementation requires the construction of finite-range unidirectionally invisible potentials with tunable reflection amplitude for the direction from which they are visible. Indeed it is sufficient to construct a tunable right-invisible potential, because its complex-conjugate will be left-invisible \cite{pra-2013a}. In view of this observation, we confine our discussion to the study of right-invisible potentials unless otherwise is clear from the context.

The finite-range exponential potentials (\ref{exp-pot}) with $k=k_0=m\pi/L$ are not suitable for the use in the above-mentioned local inverse scattering procedure, because to generate sizable values for their left reflection coefficient, one needs to take large values of $m$ for which the approximate unidirectional invisibility of the potential is destroyed \cite{longhi-jpa-2011,pra-2014a}. This calls for a method of constructing finite-range potentials that support exact unidirectionally invisibility from the right and has a tunable left reflection amplitude at the wavelength for which it is right-invisible. The standard inverse scattering schemes \cite{IS} prove to be intractable for realizing this goal. This is  because unidirectionally reflectionless and invisible potentials are necessarily complex-valued \cite{pra-2013a}, and the inverse scattering theory for complex potentials \cite{IS-complex} is too complicated to produce an explicit analytic expression for finite-range complex potentials with these properties. To the best of our knowledge, the only available method that is capable of serving this purpose is the one proposed in the context of dynamical formulation of time-independent scattering \cite{ap-2014}. The unidirectionally invisible potentials that one obtains using this method are however more complicated than (\ref{exp-pot}).

The basic motivation for the present investigation is to seek whether and to what extent the above mentioned shortcoming of the potential (\ref{exp-pot}) is a consequence of setting $k=k_0$. A complete resolution of this question requires a thorough investigation of the scattering properties of this potential for arbitrary values of $k$. The dynamical formulation of scattering theory provides a convenient method of dealing with this problem, and as we show in the sequel allows for a complete characterization of the exact unidirectional as well as bidirectional invisible configurations of this potential.

The organization of the article is as follows. In Sec.~\ref{Sec2} we review the necessary ingredients of the dynamical formulation of scattering theory and prove a theorem that provides an alternative method of calculating left reflection amplitudes. In Sec.~\ref{Sec3} we offer an analytic treatment of the scattering properties of the potential (\ref{exp-pot}) with $k_0$ being an integer multiple of $\pi/L$. In Sec.~\ref{Sec4} we present our characterization of exact unidirectional and bidirectional invisibility for this potential, and in Sec.~\ref{Sec5} we address the problem of finding its spectral singularities. Sec.~\ref{Sec6} summarizes our findings and presents our concluding remarks.

\section{Time-Independent Scattering as a Dynamical Phenomenon}
\label{Sec2}

The formulation of the time-independent scattering theory that is offered in Ref.~\cite{ap-2014} relies on the observation that the transfer matrix of one-dimensional scattering theory \cite{razavy,prl-2009,sanchez}, which encodes all the information about the scattering features of a potential, may be identified with the $S$-matrix of an effective non-unitary and non-stationary two-level quantum system (in the interaction picture \cite{pra-2014a}.) More specifically, the transfer matrix $\bM$ of any finite-range potential $v$ with support $[a_-,a_+]$ has the form:
	\be
	\bM(k)=\boldsymbol{\sU}_k(k a_+),
	\label{TM=}
	\ee
where $k$ is an arbitrary wavenumber, $\boldsymbol{\sU}_k(\tau)$ is the solution of the initial-value problem:
	\begin{align}
	&i\frac{d}{d\tau}\boldsymbol{\sU}_k(\tau)=\boldsymbol{\sH}_k(\tau)\boldsymbol{\sU}_k(\tau),
	&& \boldsymbol{\sU}_k(ka_-)=\bI,
	\label{eff-sch-eq}
	\end{align}
$\boldsymbol{\sH}_k(\tau)$ is the non-Hermitian effective matrix Hamiltonian:
	\be
	\boldsymbol{\sH}_k(\tau):=\frac{v(\tau/k)}{2k^2}\left[\begin{array}{cc}
	1 & e^{-2i\tau}\\
	-e^{2i\tau}&-1\end{array}\right],
	\nn
	\ee
and $\bI$ is the $2\times 2$ identity matrix.\footnote{If $v$ is real-valued, $\boldsymbol{\sH}_k(\tau)$ is $\boldsymbol{\sigma}_3$-pseudo-Hermitian \cite{p123}, i.e.,
$\boldsymbol{\sH}_k(\tau)^\dagger=\boldsymbol{\sigma}_3\boldsymbol{\sH}_k(\tau)
\boldsymbol{\sigma}_3^{-1}$, where $\boldsymbol{\sigma}_3$ is the diagonal Pauli matrix. Otherwise it is pseudo-normal in the sense that $[\boldsymbol{\sH}_k(\tau),\boldsymbol{\sH}_k(\tau)^\sharp]=0$, where $\boldsymbol{\sH}_k(\tau)^\sharp:=\boldsymbol{\sigma}_3^{-1}\boldsymbol{\sH}_k(\tau)\boldsymbol{\sigma}_3$
is the $\boldsymbol{\sigma}_3$-pseudo-adjoint of $\boldsymbol{\sH}_k(\tau)$, \cite{ap-2014}.}

Next, we recall that the entries $M_{ij}(k)$ of $\bM(k)$ are related to the left/right reflection and transmission amplitudes\footnote{By definition, $R^{\rm l/r}(k)$ and  $T(k)$ determine the asymptotic expression for the left- and right-incident scattering solutions $\psi_k^{\rm l/r}$ of the Schr\"odinger equation, $-\psi''(x)+v(x)\psi(x)=k^2\psi(x)$, according to
	\[\psi^{\rm l}_k(x)=\left\{\begin{array}{ccc}
	e^{ikx}+R^l(k)e^{-ikx}&{\rm for}&x\to-\infty,\\
	T(k)e^{ikx}&{\rm for}&x\to\infty,\end{array}\right.~~~~
	\psi^{\rm r}_k(x)=\left\{\begin{array}{ccc}
	T(k)e^{-ikx}&{\rm for}&x\to-\infty,\\
	e^{-ikx}+R^{\rm r}(k)e^{ikx}&{\rm for}&x\to\infty.\end{array}\right.\]}, $R^{\rm l/r}(k)$ and $T(k)$, of the potential according to \cite{prl-2009}
    \be
    \begin{aligned}
    &M_{11}(k)=T(k)-\frac{R^{\rm l}(k)R^r(k)}{T(k)},~~~~
    &&M_{12}(k)=\frac{R^{\rm r}(k)}{T(k)},\\
    &M_{21}(k)=-\frac{R^{\rm l}(k)}{T(k)},
    &&M_{22}(k)=\frac{1}{T(k)}.
    \end{aligned}
    \label{M-RT}
    \ee
This suggests that we can use (\ref{TM=}) and (\ref{eff-sch-eq}) to express them also in terms of the solution of an initial-value problem. Pursuing this idea, we find \cite{ap-2014}:
	\bea
	R^{\rm l}(k)&=&-\int_{\cC} dz~\frac{S_k''(z)}{S_k(z)S_k'(z)^2},
	\label{RL=}\\
	R^{\rm r}(k)&=&\frac{S_k(z_+)}{S_k'(z_+)}-z_+,
	\label{RR=}\\
	T(k)&=&\frac{1}{S_k'(z_+)},
	\label{T=}
	\eea
where $\cC$ is the clockwise oriented curve $\{ e^{-2ikx}\,|\,x\in[a_-,a_+]\}$ in the complex plane, $z_\pm:=e^{-2ia_\pm k}$ are its endpoints, $S_k:\cC\to\C$ is the solution of the initial-value problem,
	\bea
	&&z^2S_k''(z)+\left[\frac{\check v(z)}{4k^2}\right]S_k(z)=0,~~~~~z\in\cC,
    	\label{S-eqn}\\
	&&S_k(z_-)=z_-,~~~~~~S_k'(z_-)=1,
    	\label{ini-condi}
	\eea
and
    \be
    \check v(z):=v(\mbox{\large$\frac{i\ln z}{2k}$}),
    \label{check-v}
    \ee
so that
	\be
	v(x)=\check v(e^{-2ikx}).
	\label{v=v}
	\ee

Equations~(\ref{RL=}) -- (\ref{ini-condi}) offer a simple local inverse scattering prescription \cite{ap-2014}. For example, in order to construct a potential with support $[0,L]$
that is invisible from the right for a value $k_0$ of the wavenumber $k$, we set
$a_-=0$, $a_+=L$, $k=k_0$ and try to find a twice-differentiable function $S_{k_0}$ that in addition to the initial conditions (\ref{ini-condi}), which take the form
	\be
	S_{k_0}(1)=S_{k_0}'(1)=1,
	\label{ini-2}
	\ee
satisfies
	\begin{align}
	&S_{k_0}(e^{-2ik_0L})=e^{-2ik_0L}, &&
	S'_{k_0}(e^{-2ik_0L})=1.
	\label{invis}
	\end{align}
Clearly, these ensure that $R^{\rm r}(k_0)=0$ and $T(k_0)=1$. Therefore the potential $v$ that corresponds to this solution of (\ref{S-eqn}) is right-invisible at $k=k_0$. To determine the explicit form of $v$, it suffices to substitute the chosen $S_{k_0}$ in (\ref{S-eqn}), solve for $\check v(z)$, and employ (\ref{v=v}).

Depending on the choice of the function $S_{k_0}$ that obeys (\ref{ini-2}) and (\ref{invis}), this prescription yields different right-invisible potentials. These are unidirectionally invisible provided that $R^{\rm l}(k_0)\neq 0$. A particularly interesting situation, where we can evaluate $R^{\rm l}(k_0)$ using the machinery of contour integration, is when $k_0$ is an integer multiple of $\pi/L$. In this case, Eqs.~(\ref{invis}) coincide with the initial conditions (\ref{ini-2}), $\cC$ is the contour that wraps $m$-times around the circle $|z|=1$ in the clockwise sense, and we can evaluate the integral on the right-hand side of (\ref{RL=}) by examining the zeros of $S_{k_0}$ and $S_{k_0}'$ that are encircled by $\cC$. For a particular example, see \cite{pra-2014b}.

Note also that whenever $k_0=\pi m/L$, the outcome of the above local inverse scattering procedure is a locally periodic potential. In particular if we choose $S_{k_0}$ to be a polynomial,  $v$ will be a rational function of $e^{-2ik_0 x}$ with $x\in[0,L]$. Obviously (\ref{exp-pot}) is the simplest example of such a potential, but it does not fulfil the right-invisibility condition (\ref{invis}). At first sight, this seems to conflict with the fact that for $k_0=\pi m/L$ the invisibility condition (\ref{invis}) coincides with the initial conditions (\ref{ini-2}). Therefore, if $S_{k_0}$ satisfies the latter, $v$ must necessarily be right-invisible. This argument relies on the assumption that $S_{k_0}$ is a single-valued function on $\cC$. In general, the solution of the initial-value problem defined by (\ref{S-eqn}) and (\ref{ini-condi}) may be multi-valued. Therefore the above argument is inconclusive. As we show in Sec.~\ref{Sec3} this is precisely why the potential (\ref{exp-pot}) fails to be (exactly) right-invisible for $k=k_0=\pi m/L$.

We close this section, by a discussion of an alternative method of determining $R^l(k)$ that avoids the evaluation of the integral on the right-hand side of (\ref{RL=}). This is based on the transformation property of the entries $M_{ij}(k)$ of the transfer matrix under time-reversal transformation $\cT$, namely \cite{jpa-2014c}
    \begin{align}
    &M_{11}(k)\stackrel{\cT}{\longleftrightarrow}M_{22}(k)^*,
    &&M_{12}(k)\stackrel{\cT}{\longleftrightarrow}M_{21}(k)^*.
    \label{M-trans}
    \end{align}
We can use (\ref{M-RT}) and (\ref{M-trans}) together with the fact that $v^*$ is the time-reversal of $v$ to prove the following theorem.
    \begin{itemize}
    \item[]{\bf Theorem~1} Let $v:\R\to\C$ be a scattering potential\footnote{We use the term ``scattering potential'' to refer to potentials $v$ with a sufficiently fast asymptotic decay rate so that the general solution of the Schr\"odinger equation, $-\psi''(x)+v(x)\psi(x)=k^2\psi(x)$, tend to a linear combination of the plane waves $e^{\pm ikx}$ as $x\to\pm\infty$.} with left/right reflection amplitude $R^{\rm l/r}(k)$ and transmission amplitude $T(k)$, and $R^{\rm r}_{v*}(k)$ denote the right reflection amplitude of $v^*$, where $k$ is a (real and positive) wavenumber. Then
            \be
            R^{\rm l}(k)=\frac{T(k)^2 R^{\rm r}_{v^*}(k)^*}{R^{\rm r}(k)R^{\rm r}_{v^*}(k)^*-1}.
            \label{R-T-R}
            \ee
    \item[]{\em Proof:} Let us label the transmission amplitude of $v^*$ by $T_{v^*}(k)$. Then  (\ref{M-RT}) and (\ref{M-trans}) imply
            \begin{align*}
            &T(k)-\frac{R^{\rm l}(k)R^{\rm r}(k)}{T(k)}=\frac{1}{T_{v^*}(k)^*},
            && -\frac{R^{\rm l}(k)}{T(k)}=\frac{R^{\rm r}_{v^*}(k)^*}{T_{v^*}(k)^*}.
            \end{align*}
        Eliminating $T_{v^*}(k)^*$ in these two equations and solving for $R^{\rm l}(k)$ give (\ref{R-T-R}).~$\square$
    \end{itemize}
    \begin{itemize}
    \item[]{\bf Corollary~1} Let $\mu_\pm$ be a pair of positive real numbers and $v:\R\to\C$ be a scattering potential such that $e^{\pm\mu_\pm x}|v(x)|$ remains bounded as $x\to\pm\infty$. Then $R^{\rm l/r}(k)=0$ if and only if $R^{\rm r/l}_{v^*}(k)=0$.
    \item[]{\em Proof:} According to (\ref{R-T-R}), $R^{\rm r}_{v^*}(k)=0$ implies $R^{\rm l}(k)=0$. The converse holds because for this type of rapidly decaying potentials, $M_{ij}$ are holomorphic functions on the strip defined by $-\mu_-<\IM(k)<\mu_+$ in the complex $k$-plane, where $\IM(k)$ is the imaginary part of $k$, \cite{IS-complex}. This in particular means that $M_{22}(k)$ has no singularities on the real $k$ axis. Therefore $T(k)\neq 0$ for real $k$. In view of (\ref{R-T-R}), this completes the proof of ``$R^{\rm l}(k)=0$ if and only if $R^{\rm r}_{v^*}(k)=0$.'' Exchanging the roles of $v$ and $v^*$ in this statement, we have ``$R^{\rm l}_{v^*}(k)=0$ if and only if $R^{\rm r}(k)=0$.''~$\square$
    \end{itemize}
Notice that for finite-range potentials the hypothesis of Corollary~1 holds for all $\mu_\pm>0$. This implies that the entries of the transfer matrix for every finite-range potential are entire functions of $k$, and as a result its reflection and transmission amplitudes are meromorphic functions of $k$. This allows us to obtain the value of $R^{\rm l/r}$ and $T$ at any wavenumber $k_\star$ by evaluating the $k\to k_\star$ limit of $R^{\rm l/r}(k)$ and $T(k)$, respectively.

\section{Scattering Properties of Truncated $e^{-2ik_0x}$ Potentials}
\label{Sec3}

The potential (\ref{exp-pot}) is periodic on its support $[0,L]$ for
     \be
     k_0=\frac{m\pi}{L},~~~~~m=1,2,3,\cdots.
     \label{k0=}
     \ee
In this section we explore the scattering properties of this potential, particularly within the context of its optical realizations \cite{lin} where the coupling constant $\fz$ is related to the wavenumber $k$ and the permittivity of the optical medium $\varepsilon(x)$ according to $\fz=k^2[1-\varepsilon(0)]$.

Let us introduce the following dimensionless parameters
    \bea
    \fa&:=&\frac{\sqrt \fz}{k_0}=i\gamma\sqrt{\varepsilon(0)-1},
    \label{a=}\\
    \gamma&:=&\frac{k}{k_0}=\frac{kL}{\pi m},
    \label{g=}
    \eea
and use (\ref{coupling}) and (\ref{a=}) to express the permittivity profile associated with the potential (\ref{exp-pot}) in the form
    \be
    \varepsilon(x)=1+[\varepsilon(0)-1]e^{-2ik_0x}=1-\frac{\fa^2}{\gamma^2}\,e^{-2ik_0x},
    \label{permit-1}
    \ee
for $x\in[0,L]$, and $\varepsilon(x)=1$ for $x\notin[0,L]$.

We begin our investigation by considering the wavenumbers $k$ for which $\gamma$ is not an integer. In view of the argument given below Corollary~1, we can determine the values of the reflection and transmission amplitudes for integer values $n$ of $\gamma$ by evaluating the $\gamma\to n$ limit of their expression for non-integer $\gamma$.

According to (\ref{exp-pot}) and (\ref{check-v}), we have $\check v(z)=\fz \, z^{1/\gamma}$. Inserting this in (\ref{S-eqn}) gives
    \be
    S_k''(z)+\frac{\fa^2 z^{-2+\frac{1}{\gamma}}}{4\gamma^2}S_k(z)=0.
    \label{S-eqn-g}
    \ee
For non-integer values of $\gamma$, the solution of this equation that fulfils the initial conditions (\ref{ini-2}) has the form
    \be
    S_k(z)=\frac{-\pi\fa\,\sqrt z}{2\sin(\pi\gamma)}\left[J_{-\gamma-1}(\fa)J_\gamma(\fa \, z^{\frac{1}{2\gamma}})+J_{\gamma+1}(\fa)J_{-\gamma}(\fa\, z^{\frac{1}{2\gamma}})\right],
    \label{S=Case2}
    \ee
where $J_\nu$ is the Bessel function of the first kind \cite{ab-sh}. The presence of non-integer powers of $z$ and the Bessel functions with a non-integer index in (\ref{S=Case2}) is a clear sign that $S_k$ is multi-valued on $\cC$. To make this more transparent, we parameterize $\cC$ by $x\in[0,L]$ and label the values of $S_{k}$ and $S'_{k}$ on $\cC$ respectively by $\cS_0(k,x)$ and $\cS_1(k,x)$, i.e., set
    \begin{align}
    &\cS_0(k,x):=S_{k}(e^{-2ikx}),
    &&\cS_1(k,x):=S'_{k}(e^{-2ikx}).
    \label{exp-pot-S=02}
    \end{align}
In terms of $\cS_0$ and $\cS_1$, Eqs.~(\ref{RR=}) and (\ref{T=}) read
    \begin{align}
    &R^{\rm r}(k)=\frac{\cS_0(k,L)}{\cS_1(k,L)}-e^{-2ikL}, && T(k)=\frac{1}{\cS_1(k,L)},
    \label{RR-T=}
    \end{align}
and the initial conditions (\ref{ini-2}) become
    \begin{align}
    & \cS_0(k,0)=\cS_1(k,0)=1.
    \label{ini-3}
    \end{align}

Next, we recall the identities \cite{ab-sh}:
	\begin{align}
	&J_\nu(e^{im\pi}w)=e^{im\nu\pi}J_\nu(w),
	\label{J-analytic}\\
	&J_{\nu+1}(w)J_{-\nu}(w)+J_{\nu}(w)J_{-\nu-1}(w)=
	\frac{-2\sin(\pi\nu)}{\pi w},
	\label{id-1}\\
    &J_{\nu+1}(w)J_{1-\nu}(w)-J_{\nu-1}(w)J_{-\nu-1}(w)=
	\frac{4\nu\sin(\pi\nu)}{\pi w^2},
	\label{id-2}
	\end{align}
where $\nu$ and $w$ are respectively real and complex variables. Substituting (\ref{S=Case2})
in (\ref{exp-pot-S=02}) and using (\ref{g=}) and (\ref{J-analytic}) -- (\ref{id-2}), we find
    \bea
    \cS_0(k,L)&=&1-i\pi\,\fa\,\mu^* \,J_\gamma(\fa)J_{-\gamma-1}(\fa),
    \label{S0-gen}\\
    \cS_1(k,L)&=&1-\frac{i\pi\fa^2\mu}{2\gamma}\, J_{\gamma+1}(\fa)J_{-\gamma+1}(\fa),
    \label{S1-gen}
    \eea
where
	\be
	\mu:=\frac{1-e^{2\pi i m \gamma}}{2i\sin(\pi\gamma)}
	=\frac{1-e^{2ikL}}{2i\sin(kL/m)}.
	\label{M=}
	\ee

We can determine the right reflection and transmission amplitudes of $v$ for the wavenumber $k=\gamma k_0$ by inserting (\ref{S0-gen}) and (\ref{S1-gen}) into (\ref{RR-T=}). This gives
	\bea
	R^{\rm r}(k)&=&
    \frac{-i\pi\fa\,\mu^* J_{-\gamma-1}(\fa)J_{\gamma+1}(\fa)}{
	2\gamma-i\pi\fa^2\mu\, J_{-\gamma+1}(\fa)J_{\gamma+1}(\fa)},
	\label{RR-gen=}\\
	T(k)&=&\frac{2\gamma}{2\gamma-i\pi\fa^2\mu\, J_{-\gamma+1}(\fa)J_{\gamma+1}(\fa)},
	\label{T-gen=}
	\eea
where we have also employed (\ref{id-1}) and the recurrence relation \cite{ab-sh}:
    \be
    wJ_{\nu+1}(w)=2\nu\,J_\nu(w)-wJ_{\nu-1}(w).
    \label{rel2}
    \ee

In order to compute $R^{\rm l}(k)$, we can use (\ref{RL=}) and (\ref{S-eqn-g}) to express it as  $\frac{\fa^2}{4\gamma^2}\int_\cC z^{-2+1/\gamma} S_{k}'(z)^{-2}dz$ and try to evaluate this integral. A more convenient strategy is to use Theorem~1. This requires solving (\ref{S-eqn}) with boundary conditions (\ref{ini-2}) for $\check v(z)=\fz^*\, z^{-1/\gamma}$. The result is a multivalued function on $\cC$ that we can express using (\ref{S=Case2}) with $\gamma\to-\gamma$ and $\fa\to-\fa^*$. It turns out that we can obtain the value of this function and its derivative at $x=L$ by performing the same transformations together with $\mu\to-\mu$ on (\ref{S0-gen}) and (\ref{S1-gen}). Inserting these in (\ref{RR-T=}) and complex-conjugating the result, we obtain
    \be
    R^{\rm r}_{v^*}(k)^*=\frac{i\pi\fa\,\mu J_{-\gamma+1}(\fa) J_{\gamma-1}(\fa)}{
	2\gamma+i\pi\fa^{2}\mu^* J_{-\gamma+1}(\fa)J_{\gamma+1}(\fa)},
	\label{cRR-gen=}
	\ee	
where we have made use of (\ref{rel2}) and the following identities that apply for any real number $\nu$ and any pair of integers $n_1$ and $n_2$, \cite{ab-sh}.
	\begin{align*}
	&J_\nu(w^*)=J_\nu(w)^*, && J_{n_1-\nu}(-\fa)J_{n_2+\nu}(-\fa)=
	(-1)^{n_1+n_2}J_{n_1-\nu}(\fa)J_{n_2+\nu}(\fa).
	\end{align*}
We can determine the left reflection coefficient of the potential (\ref{exp-pot}) at the wavenumber $k=\gamma k_0$ by substituting (\ref{RR-gen=}), (\ref{T-gen=}), and (\ref{cRR-gen=})  in (\ref{R-T-R}). This leads to a rather lengthy expression that we do not include here.

Equations~(\ref{RR-gen=}), (\ref{T-gen=}), and (\ref{cRR-gen=}), that we obtained for non-integer values of $\gamma$, hold also for its integer values $n$. As we explained above, this follows from the fact that we can identify the values of $R^{\rm l/r}(k)$ and $T(k)$ at $k=nk_0$ by evaluating their $k\to n k_0$ limit. This gives
    \bea
    R^{\rm r}(nk_0)&=&\frac{-i\pi m \fa^2 J_{n+1}(\fa)^2}{
    2n-i\pi m\fa^2 J_{n-1}(\fa)J_{n+1}(\fa)},
    \label{RR-n=n}\\
    T(nk_0)&=&\frac{2n}{2n-i\pi m\fa^2 J_{n-1}(\fa)J_{n+1}(\fa)},
    \label{T-n=n}\\
    R^{\rm r}_{v^*}(nk_0)^*&=&\frac{i\pi m \fa^2 J_{n-1}(\fa)^2}{
    2n+i\pi m\fa^2 J_{n-1}(\fa)J_{n+1}(\fa)}.
    \label{TRR-n=n}
    \eea
where we have used
    \be
    \lim_{\gamma\to n}\mu=(-1)^{n+1}m,
    \label{mu-m}
    \ee
which follows from (\ref{M=}), and the identity $J_{-\ell}(w)=(-1)^\ell J_\ell(w)$,
which holds for every integer $\ell$, \cite{ab-sh}. Substituting (\ref{RR-n=n}) -- (\ref{TRR-n=n}) in (\ref{R-T-R}), we find
    \bea
    R^{\rm l}(nk_0)&=&\frac{-i\pi m \fa^2 J_{n-1}(\fa)^2}{
    2n-i\pi m\fa^2 J_{n-1}(\fa)J_{n+1}(\fa)}.
    \label{RL-n=n}
    \eea

\section{Exact Invisible Configurations}
\label{Sec4}

Equations~(\ref{RR-gen=}), (\ref{T-gen=}), and (\ref{cRR-gen=}) provide a set of conditions for the exact bidirectional and unidirectional invisibility of the class of potentials given by  (\ref{exp-pot}) and (\ref{k0=}). The analysis of these conditions turns out to be linked with the existence of common zeros of certain pairs of Bessel functions \cite{petropoulou-2003}. The work on the latter has a long history beginning with a 19th century conjecture known as Bourget's hypothesis. It states that $J_\nu$ and $J_{\nu+m}$ have no non-vanishing common zeros if both $\nu$ and $m$ are integers. Bourget's hypothesis follows as a corollary of its extension to rational values of $\nu$, \cite[\S 15.28]{watson}.

The particular question that we encounter in our investigation of the bidirectionally invisible configurations of the potential (\ref{exp-pot}) is that of the existence of non-vanishing common zeros of $J_{\nu+1}$ and $J_{1-\nu}$ for arbitrary positive real values of $\nu$. For integer and half-integer values of $\nu$ we can use Bourget's hypothesis and its above-mentioned extension to show that such zeros do not exist. For other values of $\gamma$ we do not have an answer to the question of the existence of these zeros. We investigate the consequences of conjecturing their nonexistence, i.e., assuming that the following holds.
    \begin{itemize}
    \item[]{\bf Conjecture~1} Let $\nu$ be a positive real number, then $J_{\nu+1}$ and $J_{1-\nu}$ have no non-vanishing common zeros.
    \end{itemize}
Furthermore, we employ the following result whose proof we provide in Appendix~A.
    \begin{itemize}
    \item[]{\bf Proposition~1} Let $\nu$ be a real number. Then $J_{\nu-1}$ and $J_{\nu+1}$ have no non-vanishing common zeros.
    \end{itemize}

With the help of Conjecture~1 and Proposition~1, we can use (\ref{RR-gen=}), (\ref{T-gen=}), and (\ref{cRR-gen=}) to obtain a complete characterization of the bidirectional and unidirectional invisibility of the potential (\ref{exp-pot}) with $k_0$ given by (\ref{k0=}). This is the main result of the present article which we state as the following characterization theorems.
    \begin{itemize}
    \item[]{\bf Theorem~2} Let $v$ be the potential (\ref{exp-pot}) with a nonzero real or complex coupling constant $\fz$. Suppose that $k_0$ is a positive integer multiple of $\pi/L$, and $k$ be any wavenumber. Then $v$ is bidirectionally invisible at $k$ if and only if $k$ is an integer multiple of $\pi/L$ but not of $k_0$.
    \item[]{\bf Theorem~3} Let $k_0$ be a positive integer multiple of $\pi/L$, $k$ be any wavenumber at which the potential (\ref{exp-pot}) is not bidirectionally invisible, and $\gamma:=k/k_0$. Then this potential is unidirectionally right-invisible (respectively left-invisible) at $k$ if and only if $\sqrt{\fz}/k_0$ is a zero of $J_{\gamma+1}$ (respectively $J_{-\gamma+1}$).
    \end{itemize}
We prove these theorems in Appendices~B and C. Here we suffice to mention that Conjecture~1 does not enter the proof of Theorem~3. If it happens to be false, there will be exceptional bidirectionally invisible configurations corresponding to common zeros of $J_{\pm\gamma+1}$. Therefore the bidirectional invisibility condition given in Theorem~2 will no longer be necessary. The following is a simple consequence of Theorem~3 and Eqs.~(\ref{RR-gen=}), (\ref{T-gen=}), and (\ref{cRR-gen=}).
    \begin{itemize}
    \item[]{\bf Corollary~2} Let $k_0$ be a positive integer multiple of $\pi/L$, $k$ be any wavenumber at which the potential (\ref{exp-pot}) is not bidirectionally invisible.
         Then this potential is unidirectionally invisible at $k$ if and only if it is transparent, i.e., $T(k)=1$.
    \end{itemize}

In the remainder of this section we explore the physical implications of Theorem~3.

Suppose that $J_{\pm\gamma+1}$ have no non-vanishing common zeros, and $k$ is not an integer multiple of $\pi/L$ or that it is an integer multiple of $k_0$. Then, according to Eq.~(\ref{a=}) and Theorem~3, an optical potential given by (\ref{exp-pot}) and (\ref{k0=}) can display exact unidirectional invisibility at $k$ if and only if
    \be
    \varepsilon(0)=\left\{\begin{aligned}
    &1-\gamma^{-2}\rho_{-\gamma+1}^2 &&\mbox{for left-invisibility},\\[6pt]
    &1-\gamma^{-2}\rho_{\gamma+1}^2 &&\mbox{for right-invisibility},
    \end{aligned}\right.
    \label{J2-zeros}
    \ee
where $\rho_\nu$ stands for a zero of $J_\nu$. Because $\gamma>0$ and $J_\nu$ can have non-real zeros only for $\nu<-1$, $\rho_{\gamma+1}$ is real. In light of (\ref{J2-zeros}), this implies that the exact right-invisible configurations are realized for materials with $\varepsilon(0)<1$. In particular, because in this case $\rho_{\pm\gamma+1}>\gamma$, the real and positive choices for $\rho_{\pm\gamma+1}$ correspond to certain negative-permittivity metamatials \cite{meta}.

\begin{figure}
	\begin{center}
	\includegraphics[scale=.65]{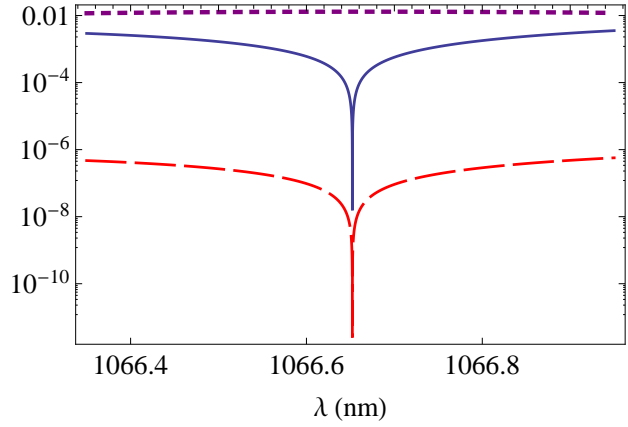}
	\caption{Plots of $|R^{\rm l}|$ (solid blue curve), $|R^{\rm r}|$ (dotted purple curve), and $|T-1|$ (dashed red curve) as a function of the wavelength, $\lambda:=2\pi/k$, for the optical potential~(\ref{exp-pot}) with $\fz=k^2[1-\varepsilon(0)]$, $k_0=\pi m/L$, and $\varepsilon(0)$, $m$, and $L$ given by (\ref{sp1}) and (\ref{sp2}). The potential displays exact unidirectional  invisibility from the left at $\lambda=1066.625~{\rm nm}$.}
	\label{fig1}
	\end{center}
	\end{figure}

We can obtain exact left-invisible configurations for ordinary material with $\varepsilon(0)>1$, provided that we take $\gamma>1$ and choose $\rho_{-\gamma+1}$ to be an imaginary zero of $J_{-\gamma+1}$. According to a theorem of Hurwitz \cite[\S 15.27]{watson}, these exist provided that $\gamma\in(2p,2p+1)$ for some positive integer $p$. In terms of $k$ this means
    \be
    2p\,k_0<k<(2p+1)k_0,~~~~~p=1,2,3,\cdots.
    \ee
Note also that, according to Hurwitz's Theorem, for each such $k$, $J_{-\gamma+1}$ has a single complex-conjugate pair of imaginary zeros which clearly correspond to a unique choice for $\varepsilon(0)$. For example, let us take $p=1$ and $\gamma=2.006200$, so that $-\gamma+1=-1.006200$. Then the imaginary zeros of $J_{-1.006200}$ are $\rho_{-1.006200}=\pm 0.157236 i$. Inserting these in (\ref{J2-zeros}), we find that our optical potential is left-invisible for $k=2.006200 k_0$ provided that
    \be
    \varepsilon(0)=1.006142617.
    \label{sp1}
    \ee
In view of (\ref{permit-1}) this gives an experimentally accessible range of permittivity modulations.\footnote{The gain/loss contrast for this configuration is of the order of $10^{-3}$.} We also choose
    \begin{align}
    &m=243, &&L=260~\mu{\rm m},
    \label{sp2}
    \end{align}
so that the wavelength associated with this value of $k$, i.e., $2\pi/k=2L/4.7m$, is $1066.625~{\rm nm}$. More importantly, we have $kL=\pi m\gamma=1531.547$. Therefore, $k$ is not an integer multiple of $\pi/L$. This observation together with Theorem~3 and the fact that for $\gamma=2.0062$ and $\fa=\pm 0.157236 i$, $J_{\gamma+1}(\fa)\neq 0$ show that the expected invisibility of the potential for this $k$ is unidirectional. Figure~\ref{fig1} shows the graphs of $|R^{\rm l/r}|$ and $|T-1|$ as functions of the wavelength $\lambda$ for the choice of the parameters given by (\ref{sp1}) and (\ref{sp2}). It clearly confirms the unidirectional invisibility of the potential for $\lambda=1066.625~{\rm nm}$.

Let us emphasize that the statement of Theorem~3 does not conflict with the existence of approximate unidirectionally invisible configurations of (\ref{exp-pot}) for integer values $n$ of $\gamma$. To describe these, we examine the behavior of $R^{\rm l/r}(nk_0)$ and $T(nk_0)$ for small values of $|\fa|$. Expanding the right-hand side of (\ref{RR-n=n}), (\ref{T-n=n}), and (\ref{RL-n=n}) in powers of $\fa$ and making use of (\ref{a=}), we have
    \bea
	R^{\rm r}(nk_0)&=&\frac{-i\pi m \fz^{n+2}}{2^{2n+3}n[(n+1)!]^2k_0^{2(n+2)}}+\cO(\fz^{n+3}),
	\label{RR=21}\\
	T(nk_0)&=&1+\frac{i\pi m\, \fz^{n+1}}{2^{2n+1}n!(n+1)!k_0^{2(n+1)}}+\cO(\fz^{n+3}),
	\label{T=21}\\
    R^{\rm l}(nk_0)&=&\frac{-i\pi m \fz^{n}}{2^{2n-1}n[(n-1)!]^2k_0^{2n}}+\cO(\fz^{n+1}).
	\label{RL=21}
	\eea
For $n=1$, which is considered in Refs.~\cite{invisible-1,lin,longhi-jpa-2011,invisible-2,jones-jpa-2012}, these relations confirm and improve the results of \cite{pra-2014a}.\footnote{There is a factor of 2 error in Eqs.~(39) and (40) of \cite{pra-2014a}. The $m$ appearing in these equations must change to $m/2$.} Clearly, they imply that for optical applications in which
	\be
	|\fz|/k_0^2=|\fa|^2= n^2|\varepsilon(0)-1|\ll 1,
	\label{pert-condi}
	\ee
and $m$ is not too large, the potential displays unidirectional invisibility (from the right) to a very good degree of accuracy. We can use the following quantities as measures of the effectiveness of the right-reflectionlessness and transparency of this potential, respectively.
    \begin{align}
    &\left|\frac{R^{\rm r}(nk_0)}{R^{\rm l}(nk_0)}\right|=
    \frac{\fz^2}{16n^2(n+1)^2 k_0^4}+\cO(\fz^3),
    &&\left|\frac{T(nk_0)-1}{R^{\rm l}(nk_0)}\right|=\frac{\fz}{4n(n+1)k_0^2}+\cO(\fz^2).
    \end{align}
These are clearly independent of $m$ and diminish as $n$ increases. Therefore, for larger values of $n$ that respect (\ref{pert-condi}), the approximate unidirectional invisibility of the potential (\ref{exp-pot}) at $k=nk_0$ has a larger domain of validity.

\section{Spectral Singularities}
\label{Sec5}

We can use the results of Sec.~3 to study the spectral singularities \cite{prl-2009} of the potential (\ref{exp-pot}) with $k_0$ given by (\ref{k0=}). For $m=1$, these have been considered in \cite{sinha} using a standard method of calculating the reflection and transmission amplitudes.

Because spectral singularities are given by the real zeros of $M_{22}$ or equivalently real poles of $T$, we can use (\ref{M=}) and (\ref{T-gen=}) to identify them with the values of $\fa$ and $\gamma$ for which
    \be
    \fa^2 J_{-\gamma+1}(\fa)J_{\gamma+1}(\fa)=\frac{4\gamma \sin(\pi\gamma)}{\pi (1-e^{2\pi i m\gamma})}.
    \label{SS-1}
    \ee
For $\gamma=n=1,2,3,\cdots$, this relation reduces to
    \be
    \fa^2 J_{n-1}(\fa)J_{n+1}(\fa)=-\frac{2in}{\pi m}.
    \label{eq-ss}
    \ee
It is easy to obtain approximate solutions of this equation for large values of $m$. Expanding the left-hand side of (\ref{eq-ss}) in powers of $\fa$ and keeping the leading order terms, we find
    \be
    \fa\approx 2\left[\frac{n! (n+1)!}{2\pi i m}\right]^{\frac{1}{2n+2}}.
    \label{af-appr}
    \ee
This relation shows that the first-order perturbation theory is most reliable for the smallest value of $n$, namely $1$. Setting $n=1$ in (\ref{af-appr}) gives $\fa^2\approx\pm 4/\sqrt{\pi i m}$. Using (\ref{a=}) and choosing the undetermined sign so that the real part of $\varepsilon(0)$ exceeds 1, we can write this equation in the form
    \[\varepsilon(0)\approx 1+\frac{4(1-i)}{2\pi m}.\]
For large values of $m$, this choice of $\varepsilon(0)$ yields an approximate spectral singularity for $k=k_0=\pi m/L$. With this input we can find the precise location of the spectral singularities by performing a numerical solution of (\ref{eq-ss}) about its approximate solutions. Table~\ref{table1} gives the values of $\fa$ and $\varepsilon(0)$ that we obtained using this method for the spectral singularities corresponding to $n=1$ and $m=100,250,500$.
    \begin{table}[!htbp]
    \begin{center}
	\begin{tabular}{|c|c|c|}
    \hline
    $m$ & $\fa$ & $\varepsilon(0)$ \\
    \hline
    100 & $0.174004 + 0.435309 i$ & $1.159217 - 0.151491 i$ \\
    \hline
    250 & $0.140574 + 0.347262 i$ & $1.100830 - 0.097632 i$ \\
    \hline
    500 & $0.119168 + 0.292458 i$ & $1.071331 - 0.069704 i$ \\
    \hline
    \end{tabular}
    \caption{Values of $\fa$ and $\varepsilon(0)$ for which the potential~(\ref{exp-pot}) has a spectral singularity at $k=k_0=\pi m/L$ with different values of $m$.}
    \label{table1}
    \end{center}
    \end{table}%

Another particular case where (\ref{SS-1}) simplifies considerably is when $\gamma$ is a half-integer; $\gamma= p+\frac{1}{2}$ with $p=0,\pm 1,\pm 2,\pm 3,\cdots$. In this case, we can  express $J_{\pm\gamma+1}$ in terms of the spherical Bessel functions $j_{\pm\gamma+1}$
with half-integer index. These admit explicit expressions involving polynomials and trigonometric functions \cite{ab-sh}. Equation~(\ref{SS-1}) has no solution for these values of $\gamma$, unless $m$ is odd. Taking $m$ odd and using the identity $J_{p+\frac{1}{2}}(w)=\sqrt{2w/\pi}\,j_p(w)$, we can write (\ref{SS-1}) as
    \be
    4\fa^3j_{p+1}(\fa)j_{-p}(\fa)=(-1)^p(2p+1).
    \label{SS-2}
    \ee
For $p=0$ which corresponds to $k=k_0/2=\pi m/2L$, this becomes
    \be
    \fa\sin(2\fa)+\cos(2\fa)=\frac{1}{2},
    \label{SS-3}
    \ee
where we have employed $j_0(w)=\sin w/w$ and $j_1(w)=\sin w/w^2-\cos w/w$, \cite{ab-sh}. To examine an optical realization of this spectral singularity, we insert (\ref{a=}) in (\ref{SS-3}) and try to solve this equation for $\varepsilon(0)$. A simple graphical inspection shows that this equation has a single positive and real solution, namely $\varepsilon(0)=4.127542$, which in view of (\ref{permit-1}) gives an unrealistically large value for the gain/loss contrast.

Other choices for $p$ in (\ref{SS-2}) and non-integer choices for $\gamma$ in (\ref{SS-1}) give similar unrealistically large values for $|\epsilon(0)-1|$. This is indeed quite expected because whenever $|\epsilon(0)-1|\ll 1$, the permittivity profile (\ref{permit-1}) corresponds to a slab with a refractive index close to unity. Therefore its boundaries are not capable of inducing sufficiently large number of internal reflections. This in turn means that the waves entering the slab do not traverse a sufficiently long optical path inside the gain region and cannot get amplified to the extend that the slab can amplify the background noise and emit purely out-going laser light which is typical of an optical spectral singularity \cite{pra-2011a}. For integer values of $\gamma$ a spectral singularity arises for $|\epsilon(0)-1|\ll 1$ provided that we take sufficiently large values for $m$. This seems to happen because in this case the waves inside the slab undergo constructive interference and have a longer optical path (because $L$ is large.)

\section{Concluding Remarks}
\label{Sec6}

The dynamical formulation of time-independent scattering theory has emerged following the observation  of a curious similarity between the composition property of the transfer matrices of scattering theory in one dimension and that of the evolution operators in quantum mechanics \cite{ap-2014}. The use of this formulation has led to a number of developments in inverse scattering \cite{pra-2014a,pra-2014b,pra-2015b,pra-2015c}, perturbative \cite{pra-2014a,pra-2015c} and semiclassical scattering \cite{jpa-2014ab}, and a recent transfer matrix formulation of scattering theory in two and three dimensions \cite{pra-2016a} which provides a powerful alternative to the standard $S$-matrix formulation of scattering theory. In this article, we proved a theorem that facilitates the application of this approach in computing the reflection and transmission amplitudes of finite-range potentials. We then used it to conduct a comprehensive study of the scattering properties of the truncated $\fz\, e^{-2ik_0x}$ potential for the cases that $k_0$ is an integer multiple of $\pi/L$. This revealed some remarkable features of this potential. In particular, it led to the discovery of exact unidirectional and bidirectional invisible configurations of this potential and allowed for their complete characterization. It also provided a simple description of the spectral singularities of this potential.

The approach pursued in this article can be applied in the study of other finite-range potentials. In general, in order to obtain exact and analytic expressions for the reflection and transmission amplitudes one has to obtain the exact solution of the initial-value problem (\ref{ini-condi}) and deal with the possible multivaluedness of its solution. These, however, do not pose any serious difficulties in the numerical implementations of this approach.

\subsection*{Acknowledgments}
I am indebted to \"Ozg\"ur M\"ustecapl{\i}o\u{g}lu for informing me of one of the references,
Eugenia Petropoulou for commenting on common zeros of Bessel function, and Keremcan Do\u{g}an, Hamed Ghaemidizicheh, and Sasan Haji-zadeh for reading the first draft of the manuscript and helping me find and correct a few errors and typos. This work has been supported by  the Scientific and Technological Research Council of Turkey (T\"UB\.{I}TAK) in the framework of the project no: 112T951, and by the Turkish Academy of Sciences (T\"UBA).

\section*{Appendix~A: Proof of Proposition~1}
A well-known property of zeros of Bessel function $J_\nu$ is that they are simple. This follows from the fact that $J_\nu$ is a holomorphic function in an open neighborhood $\cU$ of its zeros and that it satisfies~(\ref{rel2}) and the identity
    \be
    J'_\nu(z)=z J_{\nu-1}-\nu J_\nu(z).
    \label{J-prim}
    \ee
\begin{itemize}
    \item[]{\em Proof of simplicity of zeros of $J_\nu$:}
    If $z$ is a multiple zero of $J_\nu$, i.e.,  $J_{\nu}(z)=J'_{\nu}(z)=0$, (\ref{rel2}) implies $J_{\nu+2}(z)=0$. Applying this argument to $\nu+1$ gives $J_{\nu+2}(z)=0$. Repeating this, we find that $J_{\nu+m}(z)=0$ for all nonnegative integers $m$. In view of (\ref{J-prim}), this shows that all derivatives of $J_\nu$ vanish at $z$. But then, because $J_\nu$ is holomorphic on $\cU$, $J_\nu(w)=0$ for all $w\in\cU$. This contradicts the fact that zeros of holomorphic functions are isolated, and proves the simplicity of the zeros of $J_\nu$ for all real $\nu$.~$\square$ 
\end{itemize}
In the following we give a proof of Proposition~1 that makes use of this result.
\begin{itemize}
    \item[]{\em Proof of Proposition~1:} By contradiction, suppose that there is a nonzero real or complex number $z$ such that $J_{\nu-1}(z)=J_{\nu+1}(z)=0$. Then according to (\ref{rel2}), $J_{\nu}(z)=0$. Substituting this and $J_{\nu-1}(z)=0$ in (\ref{J-prim}), we have $J'_\nu(z)=0$. Because $J_\nu(z)=0$, this shows that $z$ is multiple zero of $J_\nu$. This contradicts the simplicity of zeros of $J_\nu$.~$\square$
    \end{itemize}

\section*{Appendix~B: Proof of Theorem~2}
In the following we give a proof of Theorem~2 that relies on Conjecture~1.
    \begin{itemize}
     \item[]{\em Proof of Theorem~2:} First we note that, according to (\ref{M=}), $\mu=0$ if and only if $k$ is an integer multiple of $\pi/L$ but not of $k_0$. Therefore in order to  prove this theorem it is sufficient to show that $\mu=0$ is the necessary and sufficient condition for the bidirectional invisibility of $v$ at $k$. The latter means
             \be
             R^{\rm l}(k)=R^{\rm r}(k)=T(k)-1=0.
             \label{bidir}
             \ee
According to Corollary~1 and Eqs.~(\ref{RR-gen=}), (\ref{T-gen=}), and (\ref{cRR-gen=}), this holds if $\mu=0$. To prove the converse, suppose that $\mu\neq 0$. Then the only way in which
     (\ref{bidir}) holds is that
        \be
        J_{-\gamma-1}(\fa)J_{\gamma+1}(\fa)=
        J_{-\gamma+1}(\fa)J_{\gamma+1}(\fa)=
        J_{-\gamma+1}(\fa)J_{\gamma-1}(\fa)=0,
        \label{eqns}
        \ee
Because, according to Proposition~1, $J_{\pm\gamma+1}$ and $J_{\pm\gamma-1}$ do not have common zeros, (\ref{eqns}) implies $J_{\gamma+1}(\fa)=J_{-\gamma+1}(\fa)=0$. Because $\fa\neq 0$, this contradicts Conjecture~1. Therefore, under the hypothesis that this conjecture holds, we conclude that if $\mu\neq 0$, the potential in not bidirectionally invisible.~$\square$
     \end{itemize}

\section*{Appendix~C: Proof of Theorem~3}
In the following we give a proof of Theorem~3 that does not rely on Corollary~1.
    \begin{itemize}
    \item[]{\em Proof of Theorem~3:} Because $v$ is not bidirectionally invisible, $\mu\neq 0$. Suppose that $v$ is right-invisible at $k$. Then $T(k)-1=0\neq R^{\rm l}(k)$. Equivalently, we have $T(k)-1=0\neq R_{v^*}^{\rm l}(k)^*$, which in light of (\ref{a=}), (\ref{T-gen=}), (\ref{cRR-gen=}), and $\mu\neq 0$ imply
    \bea
    J_{-\gamma+1}(\fa)J_{\gamma+1}(\fa)&=&0,
    \label{xzx1}\\
    J_{-\gamma+1}(\fa)J_{\gamma-1}(\fa)&\neq&0,
    \label{xzx2}
    \eea
where $\fa:=\sqrt\fz/k$. Equation~(\ref{xzx2}) shows that $J_{-\gamma+1}(\fa)\neq 0$. Combining this with (\ref{xzx1}) gives $J_{\gamma+1}(\fa)=0$. Similarly if $v$ is left-invisible, $T(k)-1=0\neq R^{\rm r}(k)$. Because $\mu\neq  0$, this is equivalent to (\ref{xzx1}) and
    \be
     J_{-\gamma-1}(\fa)J_{\gamma+1}(\fa)\neq  0.
    \label{xzx3}
    \ee
Clearly (\ref{xzx1}) and (\ref{xzx3}) imply $J_{-\gamma+1}(\fa)=0$. Next, suppose that $J_{\gamma+1}(\fa)=0$ (respectively $J_{-\gamma+1}(\fa)=0$). Then Eqs.~(\ref{RR-gen=}), (\ref{T-gen=}), and (\ref{cRR-gen=}) together with Corollary~1 imply that $R^{\rm l}(k)=T(k)-1=0$ (respectively $R^{\rm l}(k)=T(k)-1=0$), i.e., $v$ is right-invisible (respectively left-invisible).~$\square$
    \end{itemize}

\ed
\begin{thebibliography}{99}

\bibitem{invisible-1} L.~Poladian, Phys.\ Rev.\ E~{\bf 54}, 2963 (1996);\\
    M.~Greenberg and M.~Orenstein, Opt.\ Lett.~{\bf 29}, 451 (2004);\\
    M.~Kulishov, J.~M.~Laniel, N.~Belanger, J.~Azana, and D.~V.~Plant, Opt.\ Exp.~{\bf 13}, 3068 (2005).

\bibitem{lin} Z.\ Lin, H.\ Ramezani, T.\ Eichelkraut, T.\ Kottos, H.\ Cao, and D.\ N.\ Christodoulides, Phys.\ Rev.\ Lett.\ {\bf 106}, 213901 (2011).

\bibitem{longhi-jpa-2011} S.~Longhi, J.~Phys.~A {\bf 44}, 485302 (2011).

\bibitem{invisible-2}
    E.~M.~Graefe and H.~F.~Jones, Phys.\ Rev.~A {\bf 84}, 013818 (2011);\\
    R. Uzdin and N.~Moiseyev, Phys.\ Rev.~A {\bf 85}, 031804 (2012).

\bibitem{jones-jpa-2012} H.~F.~Jones, J.~Phys.~A {\bf 45}, 135306 (2012).

\bibitem{invisible-3}
    A.~Regensburger, C.~Bersch, M.~A.~Miri, G.~Onishchukov, D.~N.~Christodoulides, and U.~Peschel, Nature  {\bf 488}, 167 (2012);\\
    L.~Feng, Y.-L.~Xu, W.~S.~Fegasolli, M.-H.~Lu, J.~E.~B.~Oliveira, V.~R.~Almeida, Y.-F.~Chen, and
    A.~Scherer, Nature Materials {\bf 12}, 108 (2013);\\
    X.~Yin and X.~Zhang, Nature Materials {\bf 12}, 175 (2013).

\bibitem{pra-2013a} A.~Mostafazadeh, Phys.\ Rev.\ A~{\bf 87}, 012103 (2013).

\bibitem{ap-2014} A.~Mostafazadeh, Ann.\ Phys.\ (N.Y.) {\bf 341}, 77 (2014).

\bibitem{pra-2014a} A.~Mostafazadeh, Phys.\ Rev.\ A~{\bf 89}, 012709 (2014).

\bibitem{pra-2014b} A.~Mostafazadeh, Phys.\ Rev.~A.~\textbf{90}, 023833 and 055803 (2014).

\bibitem{invisible-4}
    Y.-L.~Xu, L.~Feng, M.-H.~Lu, and Y.-F.~Chen, IEEE Photonics J.\ {\bf 6}, 0600507 (2014);\\
    Y.~Shen, X.~H.~Deng, and L.~Chen, Opt.\ Express~{\bf 22}, 32053 (2014);\\
    J.~H.~Wu, M.~Artoni, and G.~C.~La~Rocca, Phys.\ Rev.\ Lett. {\bf 113}, 123004 (2014);\\
    S.~Longhi, J.~Phys.~A {\bf 47}, 485302 (2014);\\
    M.~Turduev, M.~Botey, I.~Giden, R.~Herrero, H.~Kurt, E.~Ozbay, and K.~Staliunas, Phys.\ Rev.\ A~{\bf 91}, 023825 (2015);\\
    R.~Fleury, D.~Sounas, and A.~Alu, Nature Comm.\ {\bf 6}, 5905 (2015);\\
    S.~Ding and G.P.~Wang, J.~Appl.\ Phys.~{\bf 117}, 023104 (2015);\\
    M.~Kulishov, H.~F.\ Jones, and B.~Kress,  Opt.\ Express~{\bf 23}, 18694 (2015);\\
    D.~Sounas, R.~Fleury, and A.~Alu, Phys.\ Rev.\ Appl.~{\bf 4}, 014005 (2015);\\
    Y.~Huang, G.~Veronis, and C.~J.~Min, Opt.\ Express~{\bf 23}, 29882 (2015);\\
    L.~Jin, X.~Z.~Zhang, G.~Zhang, and Z.~Song, Scientific Rep.~{\bf 6}, 20976 (2016).

\bibitem{pra-2015a} A.~Mostafazadeh, Phys.\ Rev.~A.~\textbf{91}, 063812 (2015).

\bibitem{pra-2015b} A.~Mostafazadeh, Phys.\ Rev.~A.~\textbf{92}, 023831 (2015).

\bibitem{IS} K.~Chadan and P.~C.~Sabatier, {\em Inverse Problems in Quantum Scattering Theory,} Springer, New York, 1989.

\bibitem{IS-complex} V.~A.~Blashchak, Diff.\ Equations {\bf 4}, 1519 and 1915 (1968).

\bibitem{razavy} M.~Razavy, {\em Quantum Theory of Tunneling,} World Scientific, Singapore, 2003.

\bibitem{prl-2009} A.~Mostafazadeh, Phys.\ Rev.\ Lett.~\textbf{102}, 220402 (2009).

\bibitem{sanchez} L.\ L.\ S\'anchez-Soto, J.\ J.\ Monz\'ona, A.\ G.\ Barriuso, and J.\ F.\ Cari$\widetilde{\rm n}$ena, Phys.\ Rep.\ {\bf 513} 191 (2012).

\bibitem{p123} A.~Mostafazadeh, J.~Math.\ Phys.\ {\bf 43}, 205, 2814, and 3944 (2002).

\bibitem{jpa-2014c} A.~Mostafazadeh, J.\ Phys.~A {\bf 47}, 505303 (2014).

\bibitem{ab-sh} M.~Abramowitz and I.~A.~Stegun, {\em Handbook of Mathematical Tables},
Dover, New York, 1970.

\bibitem{petropoulou-2003} E.~N.~Petropoulou, P.~D.~Siafarikas, and I.~D.~Stabolas1,
J. Comp.\ Appl.\ Math.\ {\bf 153} 387 (2003).

\bibitem{watson} G.~N.Watson, {\em A Treatise on the Theory of Bessel Functions,} Cambridge University Press, Cambridge, 1944.

\bibitem{meta} J.~B.~Pendry, A.~J.~Holden, W.~J.~Stewart, and I.~Youngs, Phys.\
    Rev.\ Lett.~\textbf{76}, 4773 (1996);\\
    A.~Al\'u and N.~Engheta, IEEE Trans.\ Antennas Propag.~\textbf{51}, 2558 (2003);\\
    C.-H.~Hsieh, A.-H.~Lee, C.-D.~Liu, J.-L.~Han, K.-H.~Hsieh, and S.-N.~Lee, AIP Adv.\ \textbf{2}, 012127 (2012).


\bibitem{sinha} A.~Sinha and R.~Roychoudhury, J. Math.\ Phys.~{\bf 54}, 112106 (2013).

\bibitem{pra-2015c} A.~Mostafazadeh, Phys. Rev. A {\bf 92}, 023831, (2015)

\bibitem{pra-2011a} A.~Mostafazadeh, Phys.\ Rev.\ A \textbf{83}, 045801 (2011);\\
A.~Mostafazadeh and M.~Sar{\i}saman, Phys.\ Rev.\ A \textbf{91}, 043804 (2015).

\bibitem{jpa-2014ab} A.~Mostafazadeh, J.\ Phys.~A {\bf 47}, 125301 and 345302  (2014).

\bibitem{pra-2016a} F.~Loran and A.~Mostafazadeh, Phys.\ Rev.~A {\bf 93}, 042707 (2016).



\end{thebibliography}
